\documentclass{pnastwo}

\usepackage{graphicx}

\usepackage{amssymb,amsfonts,amsmath}
\usepackage{color} 

\newcommand{\ep}{\epsilon}

\newcommand{\glog}{\Lambda}

\begin{document}


\title{How multiplicity determines entropy: derivation of the maximum entropy principle for complex systems}

\author{Rudolf Hanel\affil{1}{Section for Science of Complex Systems, Medical University of Vienna, Spitalgasse 23, 1090 Vienna, Austria}, 
Stefan Thurner\affil{1}{}\affil{2}{Santa Fe Institute, 1399 Hyde Park Road, Santa Fe, NM 87501, USA}
\affil{3}{IIASA, Schlossplatz 1, 2361  Laxenburg, Austria}
\and Murray Gell-Mann\affil{2}{}}


\maketitle


\begin{article}

\begin{abstract} 
The maximum entropy principle (MEP) is a method for obtaining the most likely distribution functions of 
observables from statistical systems, by maximizing entropy under constraints. The MEP has found 
hundreds of applications in ergodic and Markovian systems in statistical mechanics, information theory, 
and statistics. For several decades there exists an ongoing controversy whether the notion of the 
maximum entropy principle can be extended in a meaningful way to non-extensive, non-ergodic, and 
complex statistical systems and processes. In this paper we start by reviewing how 
Boltzmann-Gibbs-Shannon entropy is related to multiplicities of independent random processes. We 
then show how the relaxation of independence naturally leads to the most general entropies that 
are compatible with the first three Shannon-Khinchin axioms, the $(c,d)$-entropies. We demonstrate 
that the MEP is a perfectly consistent concept for non-ergodic and complex statistical systems if their  
relative entropy can be factored into a generalized multiplicity and a constraint term. 
The problem of finding such a factorization reduces to finding an appropriate representation of relative entropy 
in a linear basis. In a particular example we show that path-dependent random processes with memory 
naturally require specific generalized entropies. The example is the first exact derivation of a generalized 
entropy from the microscopic properties of a path-dependent random process.
\end{abstract}



\dropcap{M}any statistical systems can be characterized by a macro-state for which there exist many 
micro-configurations that are compatible with it. The number of configurations associated with the macro-state is 
called the phase-space volume or {\em multiplicity}, $M$. Boltzmann entropy is the logarithm of the multiplicity,   
\begin{equation}
S_B=  k_B \log M \quad,
\label{boltzent}
\end{equation} 
and has the same properties as the thermodynamic (Clausius) entropy for systems such as the ideal gas \cite{kittel}. 
We set $k_B=1$. Boltzmann entropy scales with the degrees of freedom $f$ of the system. For example for $N$ 
non-interacting point particles in $3$ dimensions, $f(N)=3N$. Systems where $S_B$  scales with system size are 
called {\em extensive}. The entropy {\em per degree of freedom}  $s_B=\frac1fS_B$ is a system-specific constant. Many complex 
systems are {\em non-extensive}, meaning that if two initially insulated systems $A$ and $B$, with multiplicities 
$M_A$ and $M_B$ respectively, are brought into contact, the multiplicity of the combined system is $M_{A+B}< M_A M_B$. 
For such systems, which are typically strongly interacting, non-Markovian or non-ergodic, $S_B$ and the {\em effective} degrees of 
freedom $f(N)$ do no longer scale as $N$. Given the appropriate scaling for $f(N)$, the 
entropy $s_B$ is a finite and non-zero constant in the thermodynamic limit, $N\to\infty$. 

A crucial observation in statistical mechanics is that the distribution of all macro-state variables gets sharply 
peaked and narrow as system size $N$ increases. The reason behind this is that the multiplicities for particular 
macro-states grow much faster with $N$ than those for other states. In the limit $N\to\infty$ the probability of 
measuring a macro-state becomes a Dirac delta, which implies that one can replace the expectation value of a 
macro-variable  by its most likely value. This is equivalent to maximizing the entropy in Eq. (\ref{boltzent}) with 
respect to the macro-state. By maximizing entropy one identifies the ``typical'' micro-configurations compatible 
with the macro-state. This typical region of phase-space dominates all other possibilities and therefore characterizes the system. 
Probability distributions associated with these typical micro-configurations can be obtained in a 
constructive way by the {\em maximum entropy principle} (MEP), which is closely related to the 
question of finding the most likely distribution functions (histograms) for a given system.

We demonstrate the MEP in the example of coin tossing. Consider a sequence of $N$ independent outcomes of coin 
tosses, $x=(x_1,x_2,\cdots,x_N)$, where $x_i$ is either head or tail. The sequence $x$ contains $k_1$ heads and $k_2$ tails. 
The probability of finding a sequence with exactly $k_1$ heads and $k_2$ tails is
\begin{equation}
	 P(k_1,k_2|\theta_1,\theta_2) = {N \choose k_1} \theta_1^{k_1} \theta_2^{k_2} = M^{\rm bin}(k) G(k|\theta) \quad,  
 \label{Mk0}
\end{equation}
where $M^{\rm bin}(k)\equiv {N \choose k_1}$ is the binomial factor. 
We use the shorthand notation $k=(k_1,k_2)$ for the {\em histogram} of $k_1$ heads and 
$k_2$ tails, and $\theta=(\theta_1,\theta_2)$ for the marginal probabilities for throwing head or tail.
For the {\em relative frequencies} $p_i\equiv k_i/N$ we write $p=(p_1,p_2)$.  
We also refer to $\theta$ as the ``biases'' of the system. 
The probability of observing a particular sequence $x$ with histogram $k$ is given by
$G(k|\theta)\equiv \theta_1^{k_1}\theta_2^{k_2}$. It is invariant under 
permutations of the sequence $x$ since the coin tosses are independent.
All possible sequences $x$ with the same histogram $k$ have identical probabilities.
$M^{\rm bin}(k)$ is the respective {\em multiplicity}, representing 
the number of possibilities to throw exactly $k_1$ heads and $k_2$ tails. 
As a consequence Eq. (\ref{Mk0}) becomes the probability of finding the distribution 
function $p$ of relative frequencies for a given $N$. The MEP is used to find the most likely $p$. 
We denote the most likely histogram by $k^*(\theta,N)$, and the most likely relative frequencies 
by $p^*(\theta,N)=k^*(\theta,N)/N$.

We now identify the two components that are necessary for the MEP to hold. The first is that $P(k_1,k_2|\theta_1,\theta_2)$ 
in Eq. (\ref{Mk0}) factorizes into a multiplicity $M(k)$ that depends on $k$ only, and a factor $G(k|\theta)$ that depends on 
$k$ and the biases $\theta$. The second necessary component is that the multiplicity is related to an entropy expression. 
By using Stirling's formula, the multiplicity of Eq. (\ref{Mk0}) can be trivially rewritten for large $N$, 
\begin{equation}
	 M^{\rm bin}(k)={N \choose k_1} \sim e^{N [ - p_1\log(p_2) - p_2\log(p_2) ]} = e^{N S[p] }\quad, 
\end{equation}
where an entropy functional of Shannon type \cite{shannon} appears,
\begin{equation}
	S[p]=-\sum_{i=1}^{W=2} p_i\log p_i\quad. 
\label{shannon}
\end{equation}
The same arguments hold for {\em multinomial} processes with sequences $x$ of $N$ independent trials, where each trial 
$x_n$ takes one of $W$ possible outcomes \cite{wallis}. In that case the probability for finding a given histogram $k$ is 
\begin{eqnarray}
\label{multifac}
P(k|\theta) = M^{\rm mn}(k) \theta_1^{k_1} \theta_2^{k_2} ... \theta_W^{k_W} = M^{\rm mn}(k) G(k|\theta)  \quad, \\ 
	{\rm with } \quad  M^{\rm mn}(k)=\frac{N!}{k_1! k_2! \cdots k_W!} \sim  e^{N S[p] } \quad. \nonumber 
\end{eqnarray}
$M^{\rm mn}(k)$ is the multinomial factor and $S[p]=-\sum_{i=1}^{W} p_i\log(p_i)$. Asymptotically 
$S[p] = \lim_{N\to\infty}\frac{1}{N}\log M^{\rm mn}(k)$ holds.
Extremizing Eq. (\ref{multifac}) for fixed $N$
with respect to $k$ yields the most likely histogram, $k^*$. Taking logarithms on both sides of 
Eq. (\ref{multifac}) gives  
\begin{equation} 
\underbrace{\frac1N \log P(k|\theta)}_{ -{\rm relative\ entropy}}=
\underbrace{\frac1N \log M^{\rm mn}(k)}_{ S[p]}+
\underbrace{\frac1N \log G(k|\theta)}_{ -{\rm cross\ entropy}}\quad.
\label{logsplit}
\end{equation} 
Obviously, extremizing Eq. (\ref{logsplit}) leads to the same histogram $k^*$. 
The term $-\frac1N \log P(k|\theta)$ in Eq. (\ref{logsplit}) 
is sometimes called {\em relative entropy} or Kullback-Leibler divergence \cite{kullbackleibler1951}. 
We identify the first term on the right hand side of Eq. (\ref{logsplit}) with Shannon entropy $S[p]$, the second term 
is the so-called {\em cross-entropy} $-\frac1N \log G(k = pN |\theta) = -\sum_i p_i\log \theta_i$. 
Equation (\ref{logsplit}) states that the cross entropy is equal to entropy plus the relative entropy.
The constraints of the MEP are related to the cross entropy.
For example, let the marginal probabilities $\theta_i$ be given by the so-called Boltzmann factor, 
$\theta_i=\exp(-\alpha-\beta \ep_i)$, for the ``energy levels'' $\ep_i$, where $\beta$ is the inverse temperature, and 
$\alpha$ the normalization constant. Inserting the Boltzmann factor into the cross-entropy, Eq. (\ref{logsplit}) becomes 
\begin{equation} 
	\frac1N \log P(k|\theta)= S[p]-\alpha \sum_i p_i-\beta\sum_i p_i\ep_i \quad, 
\label{logsplitX}
\end{equation} 
which is the MEP in its usual form, where Shannon entropy gets maximized under linear constraints. $\alpha$ and $\beta$ 
are the Lagrangian multipliers for the normalization, and the ``energy'' constraint $\sum_i p_i\ep_i=U$, respectively. 
Note that in Eq. (\ref{logsplit}) we used $f(N)=N$ to scale $\log M^{\rm mn}(k)$. Any other nonlinear $f(N)$ would yield 
nonsensical results in the limit of $S[p]$, either $0$ or $\infty$. Comparing $S[p] = \lim_{N\to\infty}\frac{1}{N}\log M^{\rm mn}(k)$ 
with Eq. (\ref{boltzent}) shows that indeed, up to a constant multiplicative factor,  $s_B=S[p]$. This means  that the 
Boltzmann entropy per degree of freedom of a (uncorrelated) multinomial process is given by a Shannon type entropy functional. 
Many systems that are non-ergodic, strongly correlated, or have long memory will not be of multinomial type, 
implying that $\hat P(x|\theta)$ is not invariant under permutations of a sequence $x$. 
For this situation it is not {\em a priori} evident if a factorization of $P(k|\theta)$ 
into a $\theta$-independent multiplicity and a $\theta$-dependent term, 
as in Eq. (\ref{multifac}), is possible.
Under which conditions such a factorization is both feasible and meaningful is discussed in the next section.

\section{When does a MEP exist?}

The Shannon-Khinchin (SK) axioms\footnote{
\label{foo_shannon}
	Shannon-Khinchin axioms:
	(SK1) Entropy is a continuous function of the probabilities $p_i$ only, and should not explicitly depend on any other parameters. 
	(SK2) Entropy is maximal for the equi-distribution $p_i=1/W$.
	(SK3) Adding a state $W+1$ to a system with $p_{W+1}=0$ does not change the entropy of the system. 
	(SK4) Entropy of a system composed of 2 sub-systems $A$ and $B$, is $S(A+B)=S(A)+S(B |A)$.	
} 
\cite{shannon,Khinchin} state requirements that must be fulfilled by any entropy. For ergodic systems all four axioms hold. For 
non-ergodic ones the composition axiom (SK4) is explicitly violated, and only the first three (SK1-SK3) hold. If all four axioms hold 
the entropy is uniquely determined to be Shannon's; if only the first three axioms hold, the entropy is given by the $(c,d)$-entropy 
\cite{HTclassification,HTextensive}. The SK axioms were formulated in the context of information theory but are also sensible for 
many physical and complex systems. 

The first Shannon-Khinchin axiom (SK1) states that entropy depends on the probabilities $p_i$ only. 
Multiplicity depends on the histogram $k=pN$ only, and must not depend on other parameters. 
Up to a $N$-dependent scaling factor the entropy is the logarithm of multiplicity. 
The scaling factor $f(N)$ removes this remaining $N$-dependence from entropy, so that SK1 is asymptotically fulfilled. 
In fact SK1 ensures that the factorization $P(k|\theta)=M(k)G(k|\theta)$ 
into a $\theta$-independent {\em characteristic} multiplicity $M(k)$, and a $\theta$-dependent {\em characteristic} probability $G(k|\theta)$,
is not arbitrary.

For systems that are not of multinomial nature, we proceed as before: to obtain the most likely distribution function we try to find  
$k=k^*(\theta,N)$ that maximizes $P(k|\theta)$ for a given $N$. 
We denote the generalized relative entropy by
\begin{equation}
	D(p|\theta)=-\frac1{f(N)}\log P(k|\theta) \quad .
\label{genrelent}
\end{equation}
Note that whenever an equation relates terms containing $k$ with terms containing $p$, we always assume $p=k/N$. The 
maximal distribution $ p^* \equiv k^*/N$  therefore minimizes $D(p|\theta)$, and is obtained by solving 
\begin{equation}
	0=\frac{\partial}{\partial p_i}\left(D(p|\theta)-\alpha\left( \sum_{j=1}^W p_i-1\right)\right)
\label{minrelent}
\end{equation}
for all $i=1,2,\cdots,W$. $\alpha$ is the Lagrange multiplier for normalization of $p$.

The histogram $k=(k_1,k_2,\cdots,k_W)$ can be seen as a vector in a $W$-dimensional space. Let $e_i$ be a $W$-dimensional 
vector whose $i$'th component is $1$, and all the others are $0$. With this notation the derivative  in Eq. (\ref{minrelent}) can be 
expressed asymptotically as
\begin{equation}
	\frac{\partial}{\partial p_i}D(p|\theta) \sim \frac{N}{f(N)}\log \frac{P(k-e_i|\theta)}{ P(k|\theta)} \equiv \frac{N}{f(N)} v_i(k|\theta) \, ,
\label{minrelent2}
\end{equation}
where we write $v_i(k|\theta)$ for the log-term. We interpret $v_i(k|\theta)$ as the $i$'th component of a vector 
$v(k|\theta)\in\mathbb{R}^W$. Let $b_{ji}(k)$ be the $i$'th component of the $j$'th basis vector for any given $k$, then 
$v_i(k|\theta)$ has uniquely determined coordinates $c_j(k|\theta)$, 
\begin{equation} 
	v_i(k|\theta)=\sum_{j=1}^W c_j(k|\theta) b_{ji}(k)\quad .
\label{coordinates}
\end{equation}
$v_i(k|\theta)$ has coordinates $c_j(k|\theta)$ in any basis $b_{ji}(k)$. However,  as can be easily verified 
not all bases are compatible with SK1-SK3 (see condition (i) in the theorem below). 
The problem of factorizing $P(k|\theta)$ therefore reduces to the problem of finding an appropriate basis.
For reasons that become clear below, we choose the following Ansatz for the basis
\begin{equation}
	b_{ji}(k)= \frac{\kappa_{ji}}{\gamma_T(N,k_i)}\log \frac{M_{u,T}(k-e_i) u(N) }{M_{u,T}(k) }  \quad , 
\label{Mbasis}
\end{equation}
where the functions $M_{u,T}(k)$ are so-called {\em deformed multinomial} factors, and $\kappa_{ji}$ are some appropriately chosen constants. 
$\gamma_T(N,r)=N\left[T(r/N)-T((r-1)/N)\right]$ is a factor depending on 
a continuous, monotonic, and increasing function $T$, with $T(0)=0$, and $T(1)=1$. 
$u(n)$  ($n=0,1,2,\cdots$) are positive, monotonic increasing functions on the natural numbers\footnote{Compare D. Radcliff's math 
blog http://mathblag.wordpress.com 2011/11/17/generalized-binomial-coefficients/}. 
The freedom of choosing $\kappa_{ji}$, $u$, and $T$, in this basis provides a well defined 
framework that allows to derive the conditions for the existence of a MEP. 
Deformed multinomials are based on {\em deformed} factorials that are well known in the mathematical literature 
\cite{bhargava2000,qfactorials,carlitz,Polya,ostrowski,gunji},  and are defined as
\begin{equation}
	N \, !_u\equiv \prod_{n=1}^N u(n) \quad . 
\label{genfac}
\end{equation}
For a specific choice of $u$, deformed multinomials are then defined in a general form as 
\begin{equation}
	M_{u,T}(k)\equiv  \frac{ N \, !_{u} }{ \prod_i \left \lfloor  N T \left(\frac{k_i}{N}\right) \right \rfloor \, !_{u} }\quad,
\label{genmul}
\end{equation}
$\lfloor x \rfloor$ is the largest integer less than $x$. With the basis of Eq. (\ref{Mbasis}) we can write 
\begin{eqnarray}
	\frac{ P(k-e_i|\theta) }{ P(k|\theta) } 
	&=&\prod_{j=1}^W \left(\frac{M_{u,T}(k-e_i) u(N)}{ M_{u,T}(k)}\right)^{\frac{c_j(k|\theta)}{\gamma_T(N,k_i)}\kappa_{ji}} \nonumber \\
 	&=& \prod_{j=1}^W u\left(NT\left(\frac{k_i}{N}\right)\right)^{c_j(k|\theta) \kappa_{ji}} \quad  .
\label{prodrep1}
\end{eqnarray}
Note that this can be done for {\em any} process that produces sequences $x=(x_1,x_2,\cdots,x_N)$, where $x_n$ takes one of $W$  
values.  We can now formulate the following
\\
\\
{\bf Theorem.} 
Consider the class of processes $x=\{x_n\}_{n=1}^N$, with $x_n\in\{1,\cdots,W\}$, 
parametrized by the biases $\theta$ and the number of elements $N$. 
The process produces histograms $k$ with probability $P(k|\theta)$. Let $N$ be large and 
$k^*(\theta,N)$  be the histogram that maximizes $P(k|\theta)$.
Assume that a basis of the form given in Eq. (\ref{Mbasis}) can be found, for which 
(i) $\kappa_{1i}=1$, for all $i=1,\dots,W$, and (ii) for fixed values of $N$ and $\theta$, the coordinate 
$c_1(k|\theta)$ of $v(k|\theta)$ in this basis, as defined in Eq. (\ref{coordinates}), becomes a non-zero constant 
at $k^*(\theta,N)$\footnote{
Condition (ii) means that the first derivatives of $c_1(k | \theta)$ vanish at $k=k^*$ under the condition
$\sum k_i=N$,  $N$ being constant.
}.
Under these conditions $P(k|\theta)$ factorizes, $P(k|\theta)=M_{u,T}(k)G_{u,T}(k|\theta)$, with 
\begin{equation}
	\frac{G_{u,T}(k-e_i|\theta)}{ G_{u,T}(k|\theta)}
	= \prod_{j=2}^W u\left(NT\left(\frac{k_i}{N}\right)\right)^{c_j(k|\theta) \kappa_{ji}}\quad .
\label{prodrep2}
\end{equation} 
Moreover, there exists a MEP with generalized entropy $S[p]=\frac1{f(N)}\log M_{u,T}(k)$, for some scaling function $f(N)$. 
The factors $u(.)^{c_j(k|\theta) \kappa_{ji}}$ in Eq. (\ref{prodrep2}) 
represent the constraint terms in the MEP. The solution of the MEP is 
given by $p^*=k^*/N$.
\\

The physical meaning of the theorem is that the existence of a MEP can be seen as a 
geometric property of a given process. This reduces the problem to one of  finding 
an appropriate basis that does not violate axioms SK1-SK3, and that is also convenient. 
The former is guaranteed by the theorem, the latter is achieved by using the particular choice 
of the basis in Eq. (\ref{Mbasis}).

Condition (ii) of the theorem guarantees the existence of primitive integrals $M_{u,T}(k)$ and $G_{u,T}(k|\theta)$.
If condition (i) is violated the first basis vector $b_{1i}$ of Eq. (\ref{Mbasis}) introduces a functional in $p$ that will in 
general violate the second Shannon-Khinchin axiom SK2. Conditions (i) and (ii) together determine $S[p]$ up to a 
multiplicative constant $c_1$, which can be absorbed in a normalization constant. 
$G_{u,T}$ may be difficult to construct in practice. However, for solving the MEP it is not necessary to know $G_{u,T}$ explicitly,  it is 
sufficient to know the derivatives of the logarithm for the maximization. These derivatives are obtained simply by taking 
the logarithm of Eq. (\ref{prodrep2}). For systems that are compatible with the conditions of the theorem, in analogy 
to Eq. (\ref{logsplit}), a corresponding MEP for the general case of non-multinomial processes reads
\begin{equation} 
	\underbrace{\frac1{f(N)} \log P(k|\theta)}_{ -{\rm generalized\ rel.\ ent.}}=
	\underbrace{\frac1{f(N)} \log M_{u,T}(k)}_{ {\rm generalized\  ent.} S[p]}+
	\underbrace{\frac1{f(N)} \log G_{u,T}(k|\theta)}_{ -{\rm generalized\ cross\ ent.}} .
\label{logsplit2}
\end{equation} 
$f(N)$ has to be chosen such that for large $N$ the generalized relative entropy $D(p|\theta)=-\frac1{f(N)} \log P(k|\theta)$ 
neither becomes $0$, nor diverges for large $N$. $S[p]=\frac1{f(N)} \log M_{u,T}(k)$ is the {\em generalized entropy}, 
and $C(p|\theta)=-\frac1{f(N)} \log G_{u,T}(k|\theta)$ is the {\em generalized cross-entropy}. In complete analogy to the 
multinomial case, the generalized cross entropy equals generalized entropy plus generalized relative entropy.
Note that in general the generalized cross-entropy $C(p|\theta)$ will not be linear in $p_i$. In \cite{HTG} it was shown that the first 
three Shannon-Khinchin axioms only allow two options for the constraint terms. They  can either be linear, or of the so-called ``escort'' 
type \cite{escort}, where constraints are given by specific non-linear functions in $p_i$ \cite{HTG}. No other options are allowed.  
For the escort case we have shown in \cite{HTG,HTG2} that a duality exists such that the generalized entropy $S$, in combination with the escort 
constraint, can be transformed into the {\em dual} generalized entropy $S^*$ with a linear constraint. In other words, the non-linearity 
in the constraint can literally be subtracted from the cross-entropy and added to the entropy. Compare with the notion of the 
``{\em corrector}'' discussed in \cite{topsoe}.

\section{The generalized entropy}

We can now compute the generalized entropy from Eq. (\ref{logsplit2})
\begin{eqnarray}
	&S[p]& = \lim_{N\to\infty}f(N)^{-1}\log  M_{u,T}(k) \nonumber \\
	&=& f(N)^{-1} \left[\sum_{r=1}^{N}\log u(r)-\sum_{i=1}^W\sum_{r=1}^{NT\left(k_i/N \right)}\log u(r) \right]  \nonumber \\
 	&=& \sum_{r=1}^{N}\frac1N \frac{N  \log u(r) }{f(N)} -\sum_{i=1}^W\sum_{r=1}^{NT\left(p_i\right)}\frac1N \frac{N \log u(r)}{f(N)} \nonumber  \\
 	&=& \int_0^1dy \frac{N \log u(Ny)  }{f(N)} -\sum_{i=1}^W\int_0^{T\left(p_i\right)}dy \frac{N \log u(Ny) }{f(N)} \nonumber \\
	&=& -\sum_{i=1}^W\int_0^{p_i}dz T'(z)\frac{N \log u (NT(z)) }{f(N)}      \nonumber \\
	& & + \int_0^1dz T'(z) \frac{N \log u(NT(z)) }{f(N)} \quad ,
\label{en}
\end{eqnarray}
$T'(z)$ is the derivative with respect to $z$. Further, we replace the sum over $r$ by an integral which is correct for large $N$. 
The resulting generalized entropy is clearly of trace form. In \cite{hanel07,thurner08,HTG} it was shown that the most general form 
of trace form entropy that is compatible with the first three Shannon-Khinchin axioms, is 
\begin{equation}
S[p] = - a\left[  \sum_{i=1}^W\int_0^{p_i}dz \glog(z) - \int_0^1dz \glog(z)\right] \quad, 
\label{en2}
\end{equation}
where $\glog$ is a so-called generalized logarithm, which is an increasing function with $\glog(1)=0$, $\glog'(1)=1$, compare 
\cite{HTG,HTG2}. Comparison of the last line of Eq. (\ref{en}) with Eq. (\ref{en2}) yields the  generalized logarithm 
\begin{equation}
	a\glog(z) =T'(z)\frac{N}{f(N)}\log u(NT(z)) - b \quad ,
\label{entr}
\end{equation}
with $a>0$ and $b$ constants. By taking derivatives of Eq. (\ref{entr}), first with respect to $z$, and then with respect to $N$, one 
solves the equation by separation of variables with a separation constant $\nu$. 
Setting $b=\log\lambda$ we get  
\begin{eqnarray}
	&\glog(z)&=\frac{T'(z)T(z)^\nu-T'(1)}{T''(1)+\nu T'(1)^2} \nonumber \\
	&u(N)&= \lambda^{\left(N^\nu\right)} \nonumber\\
	&f(N)&=N^{1+\nu} \nonumber\\
	&a&=\left(\frac{T''(1)}{T'(1)}+\nu T'(1)\right)\log\lambda \quad .
\label{sol}
\end{eqnarray}
By choosing $T$ and $\nu$ appropriately one can find examples for all entropies that are allowed by the first three SK axioms, 
which are the $(c,d)$-entropies \cite{HTclassification,HTextensive}. 
$(c,d)$-entropies include most trace form entropies that were suggested in the past decades as special cases. 
The expressions $f(N)$ and $u(x)$ from Eq. (\ref{sol}) can be used in 
Eqs. (\ref{minrelent}) and (\ref{prodrep1}) to finally obtain the most likely distribution from the minimal relative entropy,  
\begin{equation}
	p_i^*=T^{-1}\left(\left[   \frac{\log \lambda}{\alpha}  \sum_{j=1}^W  c_j(Np^* | \theta)  \kappa_{ji}  \right]^{-\frac1\nu}\right)\quad , 
\label{masta}
\end{equation}
which must be solved self-consistently. $T^{-1}$ is the inverse function of $T$. 
In case that only the first two basis vectors are relevant (the generalized entropy and one 
single constraint term), we get distributions of the form 
\begin{equation}
	p_i^*=T^{-1}\left(\left[ 1+\nu(\hat \alpha+\hat \beta \ep_i) \right]^{-\frac1\nu}\right)\quad , 
\label{selfconsist2}
\end{equation}
with $\hat \alpha = \frac{1}{\nu}( \frac{\log \lambda}{\alpha} c_1 -1)$, $\hat \beta = \frac{\log \lambda}{\alpha \nu}  c_2(Np^*| \theta)$. 
In a polynomial basis, specified by $\kappa_{ji}\equiv(i-1)^{j-1}$, the equally spaced ``energy levels'' are given by $\ep_i=(i-1)$. 
Note that $c_1=1$, and $c_2(p^*N | \theta)$ depends on bias terms. 

For a specific example let us specify $T(z)=z$, and $\lambda>1$. Eqs. (\ref{sol}) and (\ref{en2}) yield
\begin{equation}
	S[p]=\left(\frac{a}{Q}\right)\frac{1-\sum_{i=1}^Wp_i^Q}{Q-1}	\quad, \quad  \left[ Q \equiv 1+\nu \right], 
\label{tsall}
\end{equation}
which is the so-called Tsallis entropy \cite{Tsallis1988}. $\gamma_T(N,r)=1$ for this choice of $T$. 
Any other choice of $T$ leads to $(c,d)$-entropies. Assuming that the basis 
has two relevant components and using the same $\kappa_{ji}$ as above, the derivative of the constraint term in the example is 
obtained from Eq. (\ref{prodrep2}),  
\begin{equation}
	\frac{d}{d p_i} \log G_{u,T}(pN | \theta)  = \log \lambda c_2(pN | \theta) (i-1) p_i^{\nu} \quad . 
\label{constr}
\end{equation}
This constraint term is obviously non-linear in $p_i$, and is therefor of escort type. Here the expression $\ep_i=(i-1)$ plays the role of equi-distant energy 
levels. The example shows explicitly that finding the most likely distribution function $p^*$ by maximization of $P(k |\theta)$ 
(minimization of relative entropy) is equivalent to maximizing the generalized entropy of Eq. (\ref{tsall}) under a non-linear constraint 
term, $\sim \hat\beta (\sum_i  \ep_i p_i^Q-U)$. In \cite{HTG} it was shown that a duality exists that allows us to obtain exactly the same 
result for $p^*$, when maximizing the dual entropy of Eq. (\ref{tsall}), given by $S^*= \left(\frac{a}{Q}\right)\frac{1-\sum_{i=1}^Wp_i^{2-Q}}{1-Q}$, 
under the {\em linear} constraint, $\beta(\sum_i  \ep_i p_i-U)$. 

\section{Example: MEP for path-dependent random processes}

We now show that there exist path-dependent stochastic processes that are out-of-equilibrium, and whose time-dependent distribution 
functions can be predicted by the MEP, using the appropriate, system-specific generalized entropy. We consider processes that produce 
sequences $x$ that increase in length at every step. At a given time the sequence is $x=(x_1,x_2,\cdots,x_N)$. At the next time 
step a new element $x_{N+1}$ will be added. All elements take one of $W$ different values, $x_i\in\{1,2,\cdots,W\}$. The system is 
path-dependent, meaning that for a sequence $x$ of length $N$ the probability $p(i|k,\theta)$ for producing $x_{N+1}=i$ depends on 
the histogram $k$ and the biases $\theta$ only. For such processes the probability to find a given histogram, $P(k|\theta)$ can be defined 
recursively by
\begin{equation}
	P(k|\theta)=\sum_{i=1}^W p(i|k-e_i,\theta)P(k-e_i|\theta)\quad .
\label{recursiveP}
\end{equation}
For a particular example let the process have the transition probability
\begin{equation}
	 p(i|k,\theta)=\frac{\theta_i }{Z(k)} \prod_{j=i+1}^W g(k_j) \quad  {\rm with} \quad g(y)=\lambda^{(y^\nu)}\quad  ,
	\label{examplecprob}
\end{equation}
where $Z(k)$ is a normalization constant, and $\lambda>0$. Let us further fix $\theta_i=1/W$. 
Note that fixing the biases $\theta$ in multinomial systems means that as  $N$ gets large one obtains $p_i^*(\theta,N)=\theta_i$, for all $i$.
Obviously $p^*$ approaches a steady state and $N$ becomes an irrelevant degree of freedom in the sense that changing $N$ 
will not change $p^*$. Fixing all $\theta_i$ asymptotically determines $p^*$ completely and leaves no room for any further constraint. 
For path-dependent processes the situation can be very different.
For example, the relative frequencies $p^*(\theta,N)$ of the process defined in Eq. (\ref{examplecprob}) never reach a steady state as $N$ 
gets larger\footnote{One can show that for such systems the inverse temperature $c_2$ approximately grows (sub) logarithmically with $N$.}.
Here, fixing $\theta$ for all $i$ still allows $p^*(\theta,N)$ to evolve with growing $N$, such that one degree of freedom remains that can 
be fixed by an additional  constraint\footnote{Additional constraints may become necessary for intermediate ranges of $N$, 
where some coordinates $c_j$ that need to vanish asymptotically (in the appropriately chosen basis) are not yet sufficiently small.}. 
The process defined in Eq. (\ref{examplecprob}) is a path-dependent, $W$-dimensional random walk that gets more and more 
persistent as the sequence gets longer. This means that in the beginning of the process all states are equi-probable ($\theta_i=1/W$). 
With every realization of state $i$ in the process, all states $j<i$ become more probable in a self-similar way, and a monotonic distribution 
function of frequencies emerges as $N$ grows. The process appears to ``cool'' as it unfolds.
\begin{figure}[t]
\begin{center}
	\includegraphics[width=0.95\columnwidth]{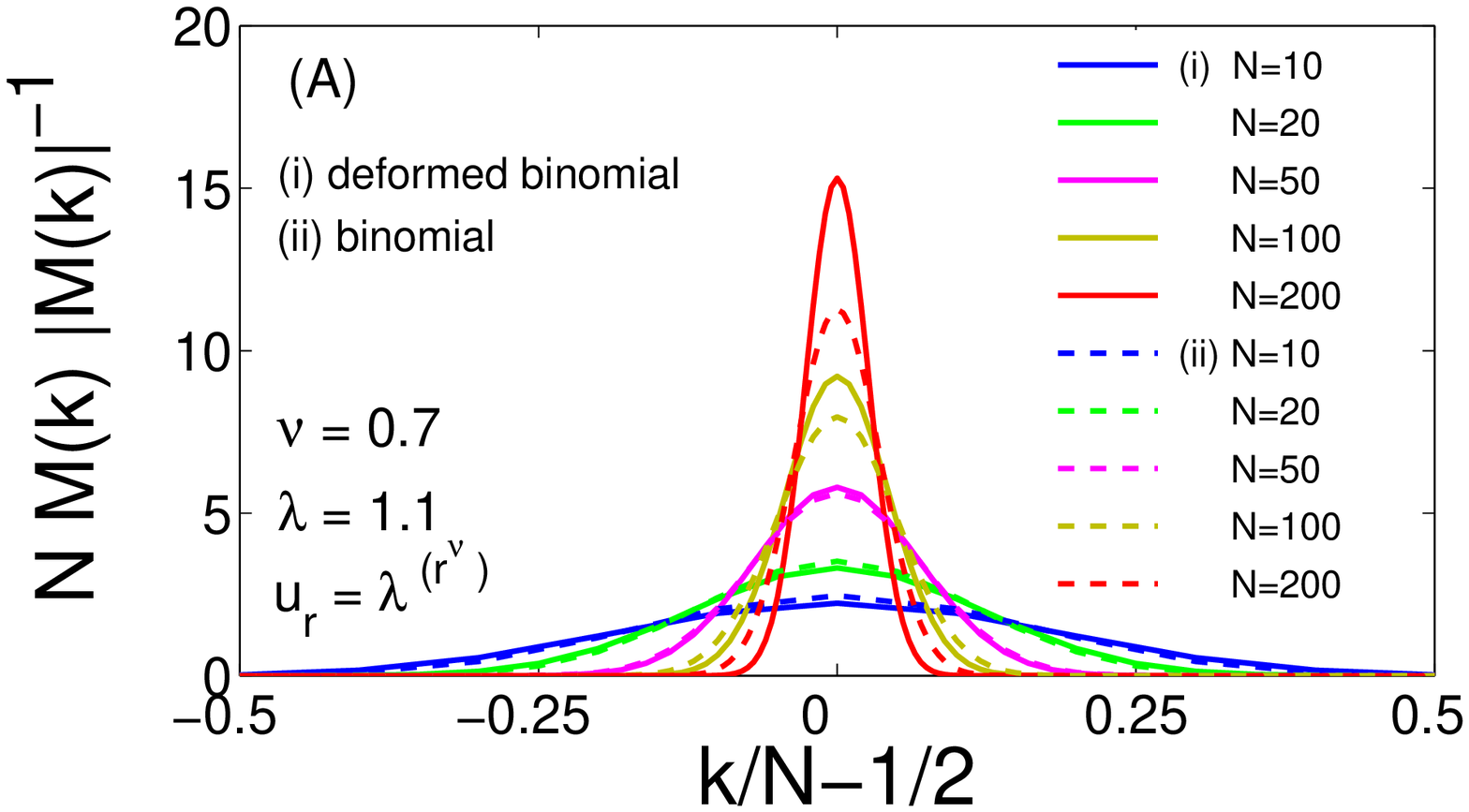}
	\includegraphics[width=0.95\columnwidth]{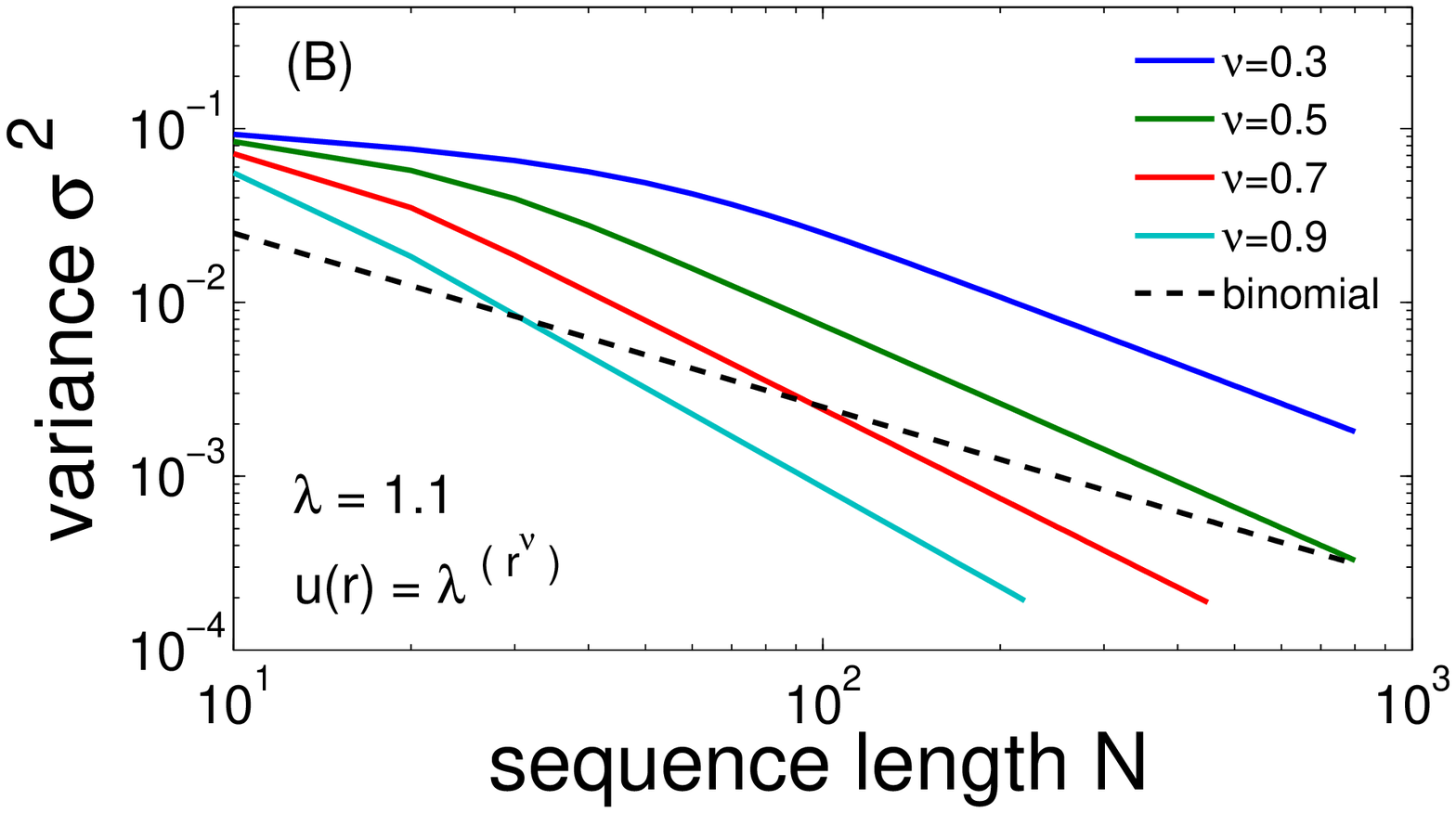} 
	\includegraphics[width=0.95\columnwidth]{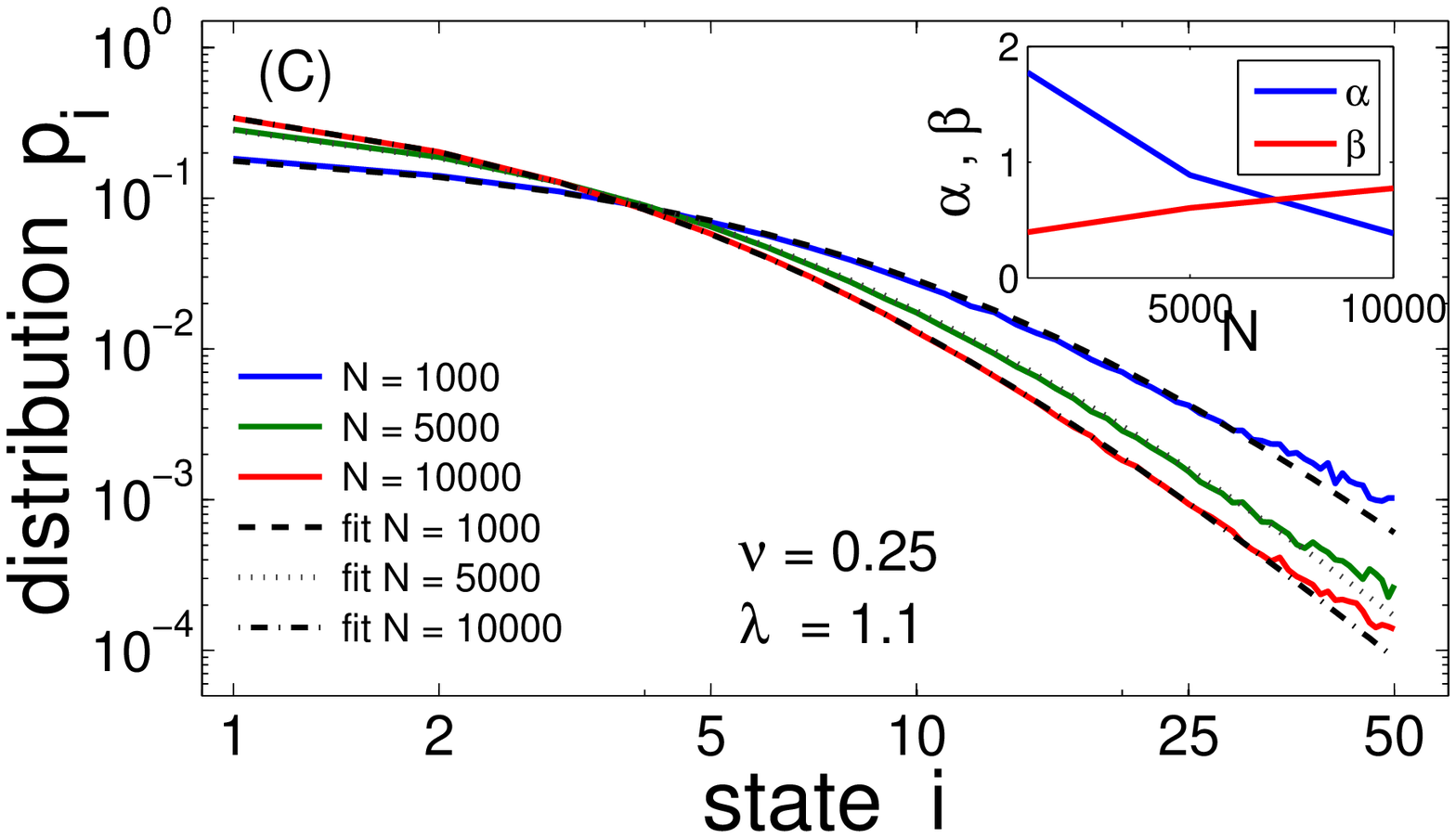}
\end{center}
\caption{
	Numerical results for the path-dependent random process determined by the deformed factorial $N!_u$ with 
	$u_r=(\lambda^{(r^\nu)}-1)/(\lambda-1)$.
	(a) Normalized generalized binomial factors $M_{u,T}(k_1,N-k_1)$ (solid lines). Distributions get narrower as $N$ increases which is necessary
	for the MEP to hold. 
   	Dotted lines show the usual binomial factor ($\nu=1$, and $\lambda\to 1$).   
	(b) Variance $\sigma^2=\sum_{k_1=0}^N M_{u,T}(k_1/N-1/2)^2$ of the normalized generalized binomial factors (solid lines), as a function 
	of sequence length $N$, for various values $\nu$,  and $\lambda=1.1$. The dashed line is the variance of the usual binomial multiplicity.
	(c) Probability distributions for the $W=50$ states $i$ from numerical realizations of  processes following Eq. (\ref{examplecprob}), with 
	$\lambda=1.1$ and $\nu=0.25$ $(Q=1.25)$ for various lengths $N$ (solid lines). Distributions follow the theoretical result from 
	Eq. (\ref{selfconsist2}). Dashed lines are $p_i=(1-(1-Q)(\alpha+\beta \ep_i))^{1/(1-Q)}$  with $\ep_i=i-1$. $\alpha$ and $\beta$ are obtained 
	from fits  to the distributions and clearly dependent on $N$ (inset). They can be used to determine $c_2$.
\label{FIGgbinomialA}
}
\end{figure}
Adequate basis vectors $b_{ji}(k)$ can be obtained with deformed multinomials $M_{u,T}(k)$ based on $u(y)=\lambda^{(y^\nu)}$, 
$T(y)=y$, and a polynomial basis for $\kappa_{ji}=(i-1)^{j-1}$. For this $u$, in Fig. \ref{FIGgbinomialA} (a) (solid lines), 
we show normalized deformed binomials for $\nu = 0.7$ and $\lambda=1.1$. Dashed lines represent the usual binomial. Clearly, generalized 
multiplicities become more peaked and narrow as $N$ increases, which is a prerequisite for the MEP to hold. In Fig. \ref{FIGgbinomialA} 
(b) the variance of deformed binomials is seen to diminish as a function of sequence length $N$ for various values of $\nu$. The 
dashed line shows the variance for the usual binomial. 
Distribution functions $p_i$ obtained for numerical simulations of sequences with $W$ states are shown in Fig. \ref{FIGgbinomialA} 
(c) for sequence lengths $N=1000$, $5000$, and $10000$ (solid lines). Averages are taken over normalized histograms from $150$ independent 
sequences that were generated with $\lambda=1.1$, and $\nu=0.25$ ($Q=1.25$). The distributions follow exactly the theoretical result from 
Eq. (\ref{selfconsist2}), confirming that a basis with 2 relevant components (one for the entropy one for a single constraint fixing $N$) is sufficient 
for the given process with $\theta_i=1/W$. 
Dashed lines are the functions suggested by the theory, $p_i=\left[1-(1-Q)(\alpha+\beta \ep_i)\right]^{1/(1-Q)}$  with $\ep_i=i-1$, where 
$\beta$ is obtained from a fit  to the empirical distribution.  $\beta$ determines $c_2$. $\alpha$ is a normalization constant. While the power exponent 
$-\frac{1}{\nu}$ does not change with $N$, the  ``inverse temperature'' $\beta$ increases with $N$ (inset), which shows that the process 
becomes more persistent as it evolves -- it ``ages''. Since $T(y)=y$, the observed distribution $p$ can also be obtained by maximizing the 
generalized entropy $S$ (Eq. (\ref{tsall})) under a non-linear constraint, or equivalently, by maximizing its dual, $S^*$ with a linear constraint, 
as discussed above. For other parameter values a basis with more than 2 components might become necessary. 
Note that the non-linear (escort) constraints can be understood as a simple consequence of the fact that the relative frequencies $p$ have to 
be normalized {\em for all} $N$. In particular the escort constraints arise from $\sum_i \frac{d}{dN} p_i^*(\theta, N)=0$, and Eq. (\ref{selfconsist2}), 
which states that $p^*$ does not change its functional shape as $\theta$ or $N$ are varied.

\section{Discussion}

We have shown that for generalized multinomial processes, where the order of the appearance of events influences the statistics of the outcome 
(path-dependence), it is possible to constructively derive an expression for their multiplicity. 
We are able to show that a MEP exists for a much wider class of processes and not only for independent multinomial processes. 
We can explicitly determine the corresponding entropic form from the transition probabilities of a system. 
We show that the logarithm of the obtained generalized multiplicity is one-to-one related to the concept of Boltzmann entropy. 
The expression for the obtained generalized entropies are no-longer of Shannon type, $-\sum_i p_i \log p_i$, but assume generalized forms, 
that are known from the entropies of superstatistics \cite{beck03,beck05} and that are compatible with the first three Shannon-Khinchin axioms 
and violate the fourth \cite{HTclassification,HTextensive,HTG}. Further, we find that generalized entropies are of trace form and are based 
on known generalized logarithms \cite{naudts,hanel07,HTG,HTG2}. 
Our findings enable us to start from a given class of correlated stochastic processes and derive their unique entropy that is needed when 
using the maximum entropy principle. We are able to determine the time dependent distribution functions of specific processes, either through  
minimization of the relative entropy or through maximization of the generalized entropy under non-linear constraints. A previously discovered 
duality allows us to obtain the same result by maximization of the dual generalized entropy under linear constraints.
Systems for which the new technology applies include out-of-equilibrium, path-dependent processes and possibly even aging systems. In an explicit 
example of a path-dependent random walk we show how the corresponding generalized entropy is derived. We implement a numerical 
realization of the process to show that the corresponding maximum entropy principle perfectly predicts the correct distribution functions as the system 
``ages'' in the sense that it becomes more persistent as it evolves. Systems of this kind often never reach equilibrium as $N\to\infty$. 

\begin{acknowledgments}
R.H. and S.T. thank the Santa Fe Institute for hospitality. M. G.-M. is glad to acknowledge the generous support of 
Insight Venture Partners and the Bryan J. and June B. Zwan Foundation.
\end{acknowledgments}


\end{article}

\begin{thebibliography}{99}

\bibitem{kittel}
C. Kittel, 
Elementary statistical physics, (Wiley 1958).

\bibitem{shannon}
C.E. Shannon,
The Bell System Technical Journal {\bf 27}, 379-423 and 623-656 (1948).

\bibitem{wallis}
E.T. Jaynes,  
{\em Probability Theory: The Logic of Science}, 
(Cambridge University Press, p. 351-355, 2003). 

\bibitem{kullbackleibler1951}
S. Kullback, R. A. Leibler,
Annals of Mathematical Statistics {\bf 22}, 79-86 (1951).

\bibitem{Khinchin}
A.I. Khinchin, 
 {\em Mathematical foundations of information theory},
(Dover Publ., New York, 1957).

\bibitem{HTclassification} 
R. Hanel, S. Thurner,
Euro. Phys. Lett. {\bf 93}, 20006 (2011). 

\bibitem{HTextensive}
R. Hanel, S. Thurner,
Euro. Phys. Lett. {\bf 96}, 50003 (2011). 

\bibitem{bhargava2000} 
M. Bhargava,
The American Mathematical Monthly, {\bf 107} 783-799 (2000). 

\bibitem{qfactorials}
F.H. Jackson,
Quart. J. Pure and Appl. Math. {\bf 41} 193-203 (1910).

\bibitem{carlitz}
L. Carlitz,
Trans. Amer. Math. Soc. {\bf 43}, 167-182 (1938).

\bibitem{Polya} 
G. Polya,
J. Reine Angew. Math. {\bf 149}, 97-116 (1919).
 
\bibitem{ostrowski} 
A. Ostrowski,  
J. Reine Angew. Math. {\bf 149}, 117-124 (1919). 
 
\bibitem{gunji}
H. Gunji and D.L. McQuillan,
J. Number Theory {\bf 2}, 207-222 (1970). 

\bibitem{HTG} 
R. Hanel, S. Thurner, M. Gell-Mann, 
PNAS {\bf 108}, 6390-6394 (2011). 

\bibitem{escort}
C. Beck, F. Schl\"ogl, 
Themodynamics of chaotic systems, 
(Cambridge University Press, 1995). 

\bibitem{HTG2} 
R. Hanel, S. Thurner, M. Gell-Mann, 
PNAS {\bf 109},  19151-19154 (2012). 

\bibitem{topsoe}
F. Topsoe,
in {\em Complexity, Metastability and Nonextensivity}, 
AIP {\bf 965}, 104-113 (2007).

\bibitem{hanel07} 
R. Hanel, S. Thurner, 
Physica A  {\bf 380},  109-114 (2007). 

\bibitem{thurner08} 
S. Thurner, R. Hanel, 
in {\em Complexity, Metastability and Nonextensivity}, 
AIP {\bf 965}, 68-75 (2007).

\bibitem{Tsallis1988}
C. Tsallis, 
J. Stat. Phys. {\bf 52} 479 (1988).

\bibitem{beck03}
C. Beck, E.D.G Cohen,  
Physica A {\bf 322}, 267-275 (2003).

\bibitem{beck05} 
C. Beck, E.G.D. Cohen, H.L. Swinney, 
Phys. Rev. E {\bf 72}, 026304 (2005).

\bibitem{naudts} 
J. Naudts, 
Physica A {\bf 316} 323-334 (2002).

\end{thebibliography}
\end{document}